\pdfoutput=1
\documentclass[prb,twocolumn]{revtex4}%,draft
\usepackage{amsmath}
\usepackage{amssymb}
\usepackage{color}
\usepackage{graphics}
\usepackage{graphicx}
\usepackage{dcolumn}

\def\q{\quad}

\begin{document}

\title {Anomalous Reflection Phase of Graphene Plasmons and its Influence on Resonators}

\author{ A.~Yu.~Nikitin$^{1,2}$}
\author{ T.~Low$^{3}$}
\author{ L.~Martin-Moreno$^{4}$}
\email{lmm@unizar.es}
 \affiliation{$^1$ CIC nanoGUNE Consolider, 20018 Donostia-San Sebasti$\mathrm{\acute{a}}$n, Spain\\
$^2$ Ikerbasque, Basque Foundation for Science, 48011 Bilbao, Spain\\
$^3$ IBM Thomas J. Watson Research Center, Yorktown Heights, New York 10598, USA\\
$^4$ Instituto de Ciencia de Materiales de Aragon and Departamento de Fisica de la Materia Condensada,
CSIC-Universidad de Zaragoza, E-50009 Zaragoza, Spain}

\begin{abstract}
The phase picked up by a graphene plasmon upon scattering by an abrupt edge is
commonly assumed to be $-\pi$. Here, it is demonstrated that for high plasmon momenta this reflection phase is  $\approx -3\pi/4$,
virtually independent on either chemical potential, wavelength or dielectric substrate.
This non-trivial phase arises from a complex excitation of highly evanescent modes close to the edge, which are required to satisfy the continuity of electric and magnetic fields.  A similar result for the reflection phase is expected for other two-dimensional systems supporting highly confined plasmons (very thin metal films, topological insulators, transition polaritonic layers, etc.). The knowledge of the reflection phase, combined with the phase picked up by the plasmon upon propagation, allows the estimation of resonator properties from the dispersion relation of plasmons in the infinite monolayer. \end{abstract}

%\pacs{42.25.Bs, 41.20.Jb, 42.79.Ag, 78.66.Bz}

\maketitle

{\it Introduction.-} Graphene can pave new ways for the development of nanoscale photonic and optoelectronic devices\cite{graphphot_natphot10}. Graphene plasmons \cite{Shung86,Vafek06,Wunsch06,Hwang07,Hansonw08,JablanPRB09,Engheta,KoppensNature12,BasovNature12} (GPs) are especially interesting due to their ultra-strong confinement, that may lead to a strong enhancement of  light-matter interaction\cite{GrigorenkoNP12,bludrev,NikitinPRB11,KoppensNL11,StauberPRB11}. Due to current limitations of the mobility of charge carriers in graphene samples, plasmon propagation lengths are not large compared to the free space wavelength $\lambda$. However, they can still be large compared to the graphene plasmon wavelength $\lambda_p$ which can be as small as $\lambda/100$. For this reason interesting resonant effects occur in graphene structures of deep subwavelength scales. For instance, resonant enhancement of absorption due to excitation of GPs has recently been proposed\cite{PlasmonicsNature11,NikitinPRBabs12,deAbajoPRL12}
and demonstrated\cite{NatureDiscs12,IBMNaturePhot13}.

In particular, resonant plasmonic effects have been measured in arrays of ribbons in an ample range of ribbons widths\cite{PlasmonicsNature11,IBMNaturePhot13,PRBRana13,IshikawaAPL13,BrarNL13,BrarNL14}. Often, as evident in these works, the analysis of the resonant absorption peaks has been based on the comparison with the dispersion relation of two-dimensional (2D) GPs via the relation $k_{p}=\pi/W$, where $W$ is the ``active'' part of the graphene ribbon (i.e. the region having a sufficiently high concentration of charge carriers). This simple relation between the modes inside the ribbon and 2D GPs corresponds to a Fabry-Perot cavity with ideally-reflecting walls and is justified by the high value of the GP reflection coefficient at a graphene edge\cite{JLACSnano13,ChenNL13}. Such an interpretation is however only valid when the phase shift of the GP reflected from the ribbon termination is $-\pi$.
%Observed deviations between the dispersion relation of GPs and the ``folded'' dispersion relation in graphene ribbons have been associated to the presence of a ``dead zone''\cite{IBMNaturePhot13} and inter-ribbon coupling effects\cite{PRBRana13}.

In this paper we show an effect that has been neglected up to now: the influence of the phase picked up by a GP upon reflection at a graphene edge. We will demonstrate that this is a very important contribution, which affects both the resonator frequency and the nature of the electromagnetic modes.

\begin{figure}[thb!]
\includegraphics[width=7cm]{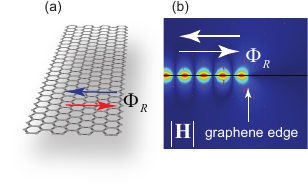}\\
\caption{(Color online) (a)  Schematic of the reflection of a graphene plasmon by an edge. A graphene plasmon propagating from left to right (shown by a red arrow) acquires a phase $\Phi_R$ and then propagates back (which is shown by a blue arrow). (b) Computed absolute value of the magnetic field $|\mathbf{H}|$ created by the interference between the incident graphene plasmon and that reflected from the edge. In the calculation graphene lies on a SiO$_2$ substrate, and the considered free-space wavelength is $12 \mu$m.  The parameters used for computing the graphene conductivity are: temperature $300$ K, Fermi level $0.35$ eV.}\label{geom}
\end{figure}

{\it Phase picked up by reflection at an edge.-} It is now well known that when a GP reaches an edge, it is reflected with virtually 100\% amplitude\cite{JLACSnano13,ChenNL13}. This effect can be attributed to the large density of states on the GP channel (much larger than the density of photonic radiating modes).

In order to investigate the phase of the reflection coefficient, let us start by analyzing an idealized system: an homogeneously-doped lossless free-standing graphene plasmon resonator (ribbon). As will be shown later, despite its simplicity this model contains very useful information.
We consider the situation, schematically represented in Fig. 1(a),  where a GP is launched (from $x<0$) at normal incidence towards a graphene edge, placed at $x=0$, and compute\cite{COMSOL} the full electromagnetic field.  In these calculations, and throughout the manuscript, graphene is modeled by its 2D conductivity $\sigma$, obtained from the random phase approximation\cite{Wunsch06,Hwang07,Falkovsky08}.
Fig. 1(b) renders the absolute value of the total magnetic field. The single-period interference pattern with regions of zero field in the plane confirm that the scattered field is essentially a back-reflected GP.

The different contributions to the magnetic field are represented in Fig 2(a), clearly showing that, beyond a transition region at very small distances to the edge, the incident and reflected wave have the same period and virtually the same amplitude, but present a phase shift.

To characterize this phase shift we define the auxiliary function $\Phi_R(x) $ through the reflected field $H_r$:
\begin{equation}\label{rloc}
H_{r}(x, z=0^+) = |H_{r}(x, z=0^+)|e^{- \imath k_p x} e^{i\Phi_R(x)}.
\end{equation}

Defined in this way, $\Phi_R(x) $ is just a re-parametrization of the phase of the reflected field but, at sufficiently long distances from the edge, it converges to the phase pick up by the GP upon the reflection
at the edge, $\Phi_R \equiv \Phi_R(x\rightarrow -\infty)$. Fig 2(b) shows the computed $\Phi_R(x) $ for several wavelengths, both for free standing graphene and for graphene on a SIO$_2$ substrate. Two regions are apparent in this figure. First, at small distances to the edge (smaller than $50-100$ nm),  $\Phi_R(x) $ varies spatially, due to the excitation of  deeply evanescent modes generated by the termination. These modes are reminiscent of the "quasi-cylindrical" waves originated by subwavelength emitters in both metals and graphene \cite{NikitinNJP09,LalanneNatPhys06} and, in this problem, are responsible for a change in phase of the order of $15\%$. Second, at larger distances, $\Phi_R(x) $ converges to a
GP reflection phase $\Phi_R \approx -0.75 \pi$. Notably, this value is virtually independent of the intrinsic properties of graphene (Fermi level, temperature, etc) and dielectric environment.

\begin{figure}[thb!]
\includegraphics[width=7cm]{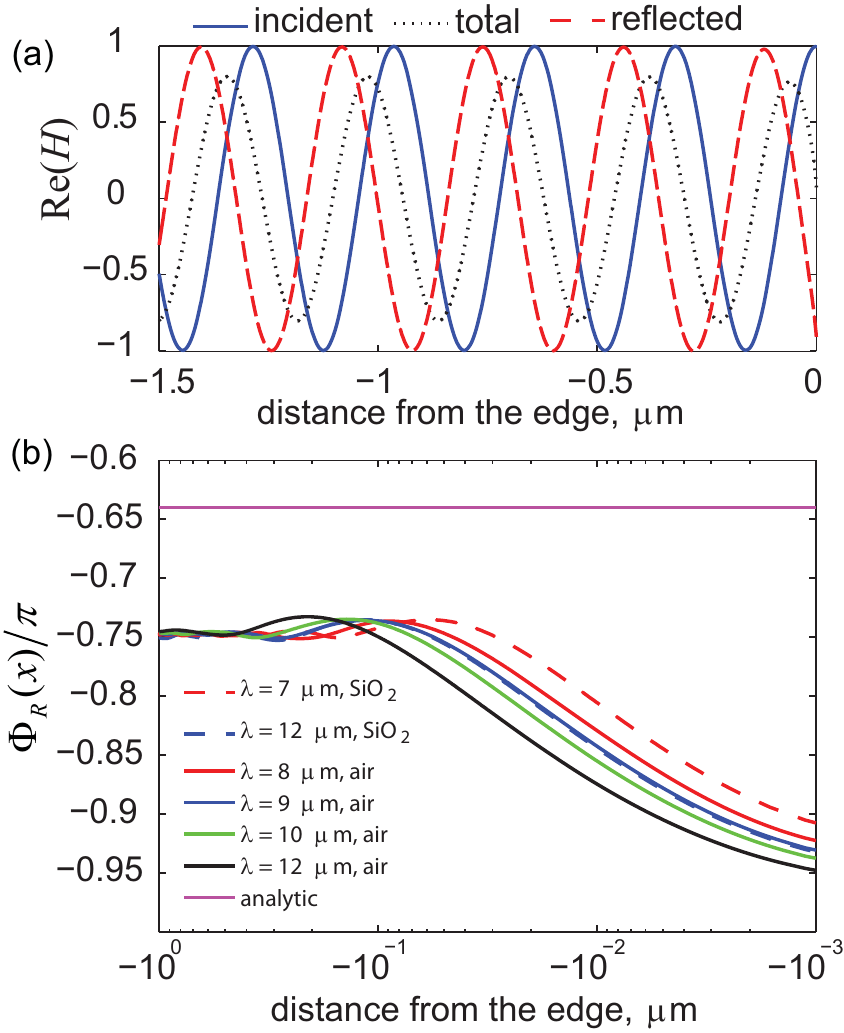}\\
\caption{(Color online) (a) Real part of the total magnetic field (dotted curve), incident GP (continuous curve), and reflected GP (dashed curve) as a function of the distance from the edge. The fields correspond to graphene on SiO$_2$ substrate at wavelength of $12 \mu$m. (b) The phase of the reflection coefficient $\arg(r)$ as a function of the distance from the edge (located at $x=0$) for different wavelength and configurations. The phase taken from the analytical result according to Eq.~\eqref{e3.2} is shown by a horizontal line.  The parameters used for graphene conductivity: temperature $300$ K, Fermi level $0.35$ eV.}\label{FiR}
\end{figure}

It is possible to obtain a good estimation of this non-trivial reflection phase, and further insight into its origin, using a simple analytical model. Following an approach that was used to study the reflection of the surface plasmons in metal slabs\cite{Gordon06}, we neglect the complex behavior of the EM field close to the edge,
and approximate the field at the graphene side by the sum of the incident and reflected GPs.
The field is expanded in vacuum ($x>0$) in its eigenstates, thus having the form:
\begin{equation}\label{e2}
H_{y}(x,z) = \begin{cases} \pm(e^{ik_px} + r_{E} e^{-ik_px})\, e^{\pm ik_{pz}z}, & x<0 \\
                                               \int_{-\infty}^{\infty} dk_z \, h(k_z) \, e^{ik_{x}x+ ik_{z}z}, &  x>0
                       \end{cases}
\end{equation}
where $r_{E}$ and $h(k_z)$ are expansion coefficients, $\pm $ stay for $z>0$ and $z<0$ respectively (the magnetic field for a GP is antisymmetric with respect to the graphene position). The wave vector components are  $k_{zp}=-g/\alpha$ (with $g=\omega/c$ and $\alpha=2\pi\sigma/c$) and  $k_{p}=\sqrt{g^2-k_{pz}^2}$,  $k_x = k_x(k_z)= \sqrt{g^2-k_z^2}$, with $\mathrm{Im}(k_{p}), \mathrm{Im}(k_{x})\geq 0$ to satisfy the radiation conditions.
 The electric field components can be found from Eqs.~\eqref{e2} using Maxwell's equations. The unknown coefficients $r$ and $h(k_z)$ are obtained by imposing continuity of the magnetic fields and the $z$-component of the electric field along the plane $x=0$ (see the Supplementary Information for the details). The final result for the reflection coefficient is
\begin{equation}\label{e3.2}
\begin{split}
r_{E} = \frac{a-1}{a+1},
  \end{split}
\end{equation}
where
\begin{equation}\label{e3.3}
\begin{split}
a = \frac{4|k_{pz}|k_p}{\pi} \int_{0}^{\infty} dk_z \frac{k_{z}^2}{k_x(|k_{pz}|^2+k_z^2)^2}.
  \end{split}
\end{equation}
This integral can be computed numerically (the results can be found in the Suppl. Mat.),
but an excellent analytical approximation can be found for
the case with large values of $k_p$ (which is the case of interest, due to strong confinement of the GP).
Noticing that the main contribution to the integral originates from $k_z\sim|k_{pz}|\gg1$,
we can approximate $k_x\simeq ik_z$, and obtain $a=-2i/\pi$, yielding a reflection phase
$\Phi_R= \mathrm{arctan}(-4 \pi / (4 + \pi^{2})) \approx - 0.64 \pi$. Thus, this analytical result
coincides with the phase extracted from full-wave calculations with $13$\% error, which is consistent with the neglected contributions from the highly confined modes in the graphene region that lead to the formation of the quasi cylindrical wave close to the edge.

In order to get a better description of experimentally observed samples, let us extend the previous discussion to situations where the carrier density, and thus the
conductivity $\sigma=\sigma(x)$, changes spatially close to the edge. In this case the reflection coefficient corresponds to the reflection from the whole inhomogeneous region, not only from the abrupt termination $x=0$. However, taking into account that GPs can follow adiabatically variations in conductivity in length scales even smaller than the GP wavelength\cite{JLACSnano13}, we can hypothesize that the GP scattering can then be split into 3 ``events'': (i) GP propagating through the inhomogeneity to the edge; (ii) reflection of the GP from the edge with the reflection coefficient $r_E$; (iii) backward propagation of the reflected GP $r$ through the
inhomogeneous region. In other words, the full reflection coefficient is simply the product of the reflection coefficient for the edge $r_E$ and phase due to the propagation of the GP along the inhomogeneity region (see details in Suppl. Mat.):
\begin{equation}\label{e4}
\begin{split}
r=r_E \, e^{i\Delta\varphi}, \q \Delta\varphi =  2i\int_{x}^0dx(k_p(x)-k_{p0}).
  \end{split}
\end{equation}
with $k_{p0} = k_{p}(-\infty)$.

\begin{figure}[thb!]
\includegraphics[width=7cm]{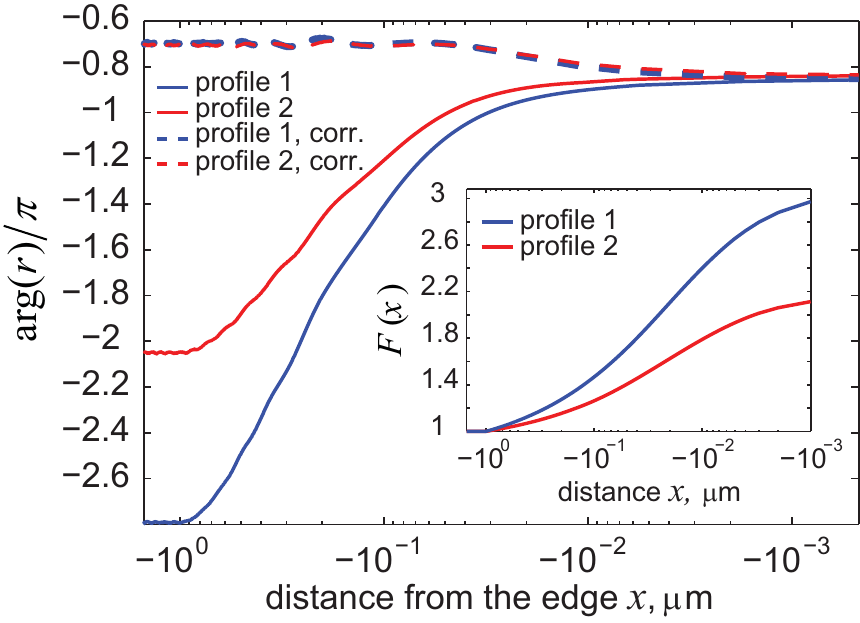}\\
\caption{(Color online) The phase of the reflection coefficient as a function of the distance from the edge (located at $x=0$) for two different profiles $\sigma(x)=\sigma_0F(x)$, with $F(x)$ shown in the inset. Both $\arg(r)$ (continuous curves) and $\arg(r)-\Delta\varphi$ (``corrected'' phase, discontinuous curves) are rendered. The parameters used for graphene are the same as in Fig. 2}\label{FiRq}
\end{figure}

In order to check the validity of this approximation, we have carried out simulations for the two different conductivity profiles, rendered in the inset to Fig 3(a).  These profiles present a $\sim1/\sqrt{x}$-like dependency for $F(x)$ (see details in the Suppl. Mat.) since similar behaviour of the Fermi level was found in electrostatically doped graphene ribbons \cite{deAbajoAPL12}. Fig 3(a) shows that the argument of the reflection coefficient is very different for both profiles (rendered by the continuous curves). However, when corrected by the optical phase picked up by propagation in the non-uniform region this argument converges to approximately $-0.7\pi$.  This phase is very close to the previously found value for the edge of a homogeneous sheet, thus confirming that, to a good approximation, the GP follows adiabatically the conductivity variation up to the edge.

\begin{figure}[thb!]
\includegraphics[width=8cm]{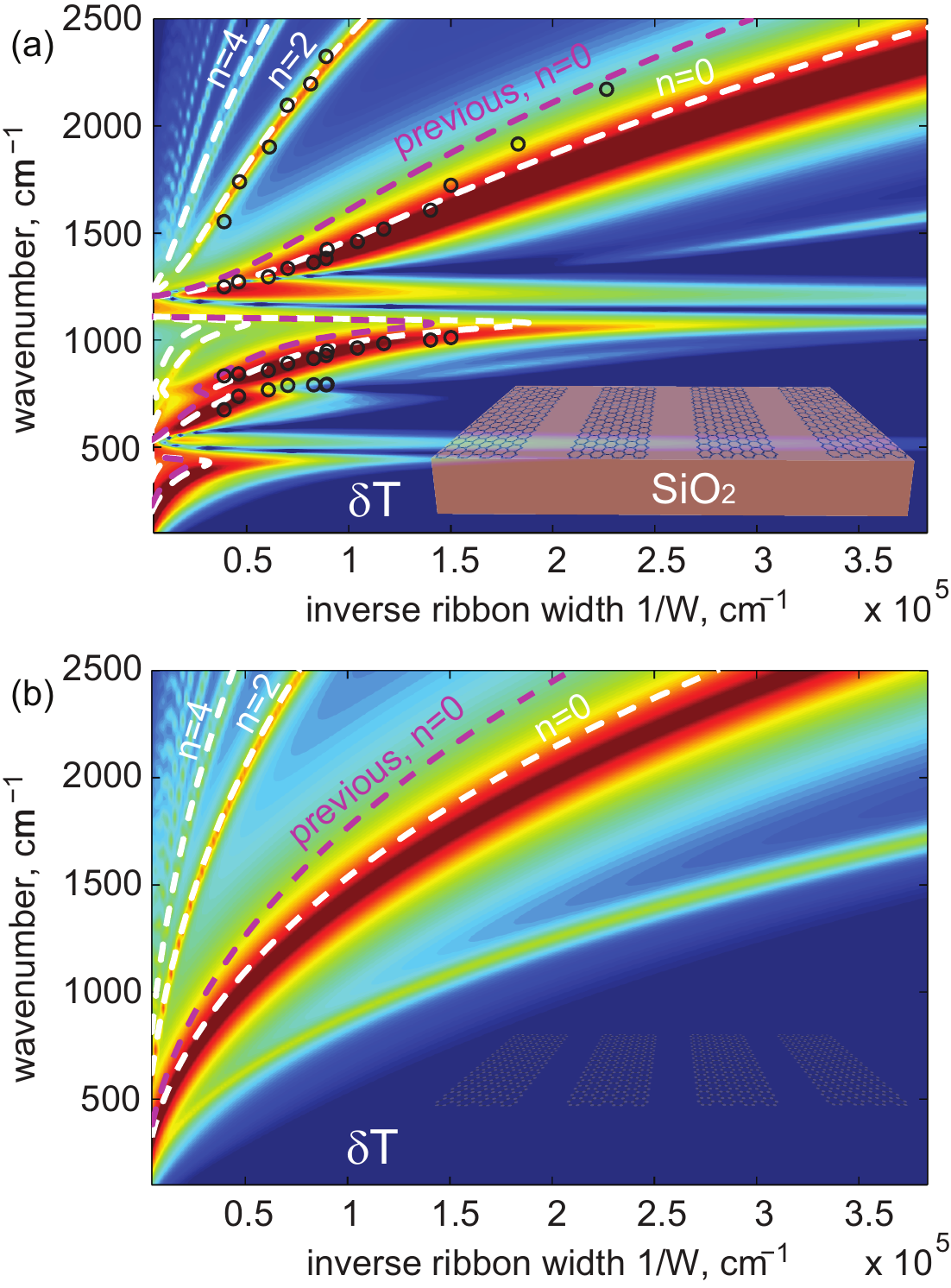}\\
\caption{(Color online) (a) Relative transmission $\delta T = |1-T/T_0|$ (with respect to the one for the graphene-free substrate) through the array of graphene ribbons on SiO$_2$ substrate as a function of the inverse ribbon width and wavenumber. The array has a period twice the ribbon's width. (b) The  same but for free-standing graphene. Circular symbols represent the experimental data. The discontinuous curves numbered by ``0,2,...'' correspond to the modes given by Eq.~\eqref{e1}. The discontinuous curve marked as ``previous, n=0'' corresponds to $\Phi_R=-\pi$ and $n=0$ in Eq.~\eqref{e1}. The parameters used for graphene conductivity: temperature $300$ K, Fermi level $0.55$ eV, relaxation time of the charge carriers is $0.1$ ps.}\label{spectra}
\end{figure}

{\it GP Modes in ribbons.-}

Several experiments have analyzed the modes in arrays of graphene ribbons, illuminated at normal incidence with an electric field pointing perpendicularly to the ribbon axis, in terms of a Fabry-Perot model, where excited plasmons propagate back and forth between the ribbon edges.
In a single homogeneous ribbon, with width $W$, the modes should satisfy
\begin{equation}\label{e1}
\begin{split}
2 k_p W + 2 \Phi_R = 2 \pi n,
  \end{split}
\end{equation}
All analysis have, to the best of our knowledge, consider that the field vanishes at the edges, thus assuming $\Phi_R =-\pi$. In this case, the fundamental mode occurs for $n=0$, leading to the commonly used expression
$k_{p} = \pi/W$. However, the existence of the non-trivial reflection phase analyzed in this manuscript modifies this picture. For instance, notice that the fundamental mode occurring for $n=0$ satisfies $k_{p} = - \Phi_{R}/W$.

In order to illustrate this point, we have carried out simulations (using modal expansion \cite{NikitinPRBabs12})  of the  transmission spectra $T$ of normal-incidence electromagnetic waves, with electric field polarized normal to the ribbons, impinging into arrays of ribbons placed on an  insulating substrate. Resonances in the relative transmission, i.e. with respect to the graphene-free substrate case, correspond to enhanced absorption due to the excitation of graphene plasmons. In order to compare with experimental data\cite{IBMNaturePhot13},  we have assumed an array
period $L=2 W$, a temperature of $300$ K, a typical carrier relaxation time of $0.1$ ps for CVD graphene, and SiO$_2$ substrate whose dielectric function is taken from Ref. \cite{IBMNaturePhot13}.
Fig. 4 (a) shows the intensity plot of the relative transmission computed for ribbon arrays with different widths. The white discontinuous lines provide the expected position of the resonances, based on Eq.~\eqref{e1},  where $k_{p}$ is given by the dispersion relation for graphene plasmons on a SiO$_2$ substrate.
%This dispersion relation can in general be easily found numerically, but in the region of large momentums $k_p\gg g\sqrt{\varepsilon_{SiO_2}}$ it reads as $k_p(\omega)=i[1+\varepsilon_{SiO_2}(\omega)]g/2\alpha(\omega)$.
 The result of the mapping for the first three even modes  $n=0,2,4$ is shown by the white discontinuous curves (modes characterized by odd integers are antisymmetric with respect to the ribbon axis and do not couple to normal incidence radiation), in reasonably good agreement with the simulations. The open circles mark the spectral position for maximum experimental reflectance taken from Ref. \cite{IBMNaturePhot13}. By assuming a Fermi level of $0.55$ eV, and an effective width of the ribbon $28$nm narrower than the nominal value, due to the presence of disorder close to the ribbon edge\cite{IBMNaturePhot13}, we obtained very good agreement between experiment and theory. Remarkably, using the anomalous reflection phase computed for graphene plasmons provides a good approximation for the spectral position of different branches, not only the $n=0$ one, for  both measured and computed resonant spectra for ribbons. Notice that the deviation of the experimental points from the theoretically predicted positions of the resonances for large values of $1/\mathrm{W}$ in Fig. 4 (a) might be attributed to the growing influence of non-homogeneously doped edges for very thin ribbons.\\
In Fig. 4b we render the computed results for a free-standing graphene sheet.  In this plot we show the full calculation for a ribbon array, together with prediction from the traditionally used expression $k =\pi/W$ (labeled ``previous, $n=0$'') and that with the anomalous reflection phase using Eq.~\eqref{e1}.  Also in this case, the traditional model captures the trends but fails to reproduce accurately the spectral position for maximum transmission. In contrast, our simple Fabry-Perot model with  non-trivial reflection phase $\Phi_{R}$ provides a good approximation to the computed resonant spectra for ribbons, specially for small momentum (larger ribbons widths). Notice that, for the considered ratio between period and ribbon width, the ribbon-ribbon interaction affects mainly the lowest frequency resonance ($n=0$) \cite{NikitinPRBabs12,PRBRana13} and provides a slight red shift (around a few percent) of the resonance peak compared to the value predicted by Eq. ~\eqref{e1}. Calculations show that already for slightly larger inter-ribbon distance ($L=3W$)  Eq.~\eqref{e1} provides a virtually exact value for the resonance spectra (see Suppl. Mat.).

{\it Conclusions.-} We have shown that when plasmon in graphene is reflected from an abrupt termination, the phase
of the reflected plasmon is largely independent on dielectric environment, wavelength and local doping provided that the GP is strongly confined. The reflection phase has a nontrivial value of $\Phi_R\approx-3\pi/4$. We reached our results using three different methods (finite elements calculation, quasi-analytical model and a fully analytical approximation). Taking this phase into account is
essential to design and analyze optical properties of graphene resonators (ribbons, discs, etc.). By appropriately remapping of the modes inside a single graphene ribbon, we have successfully fit the experimentally observed resonances in periodic arrays of graphene ribbons.
Our results can be also further extended to other two-dimensional systems supporting plasmons (like very thin metal films, topological insulators, transition polaritonic layers, etc.), where
phase of the reflection coefficient is normally, and perhaps incorrectly, assumed to be $-\pi$.

{\it Acknowledgements.-}
We express our sincere gratitude to F. Guinea for stimulating and fruitful discussions. We acknowledge
support by the Spanish Ministerio de Economia y Competitividad within project MAT2011-28581-C02. LMM acknowledges support by the EC under Graphene Flagship (contract no. CNECT-ICT-604391).

\appendix

\section{Homogeneously doped graphene}

We will solve the problem of the graphene plasmon (GP) reflection from the edge by assuming that a single mode (GP) from the region $x<0$ couples to the ``electromagnetic continuum'' in the region $x>0$.

\begin{figure}[h!]
\begin{center}
\includegraphics[width=8cm]{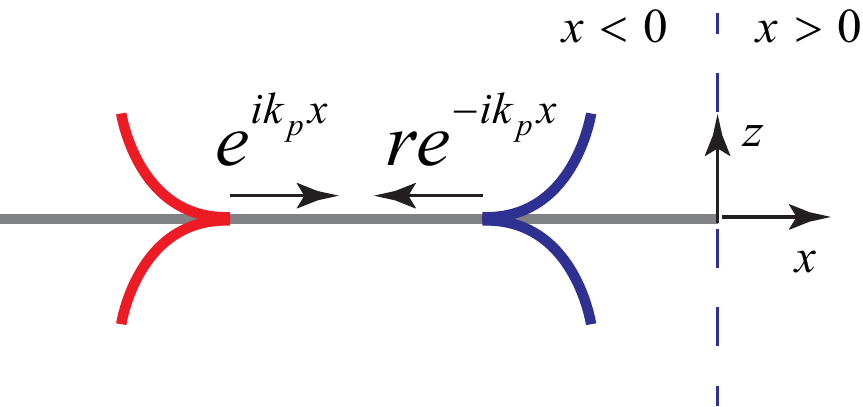}\\
\caption{The schematic of the studied system: the incident GP impinge onto the graphene termination at $x=0$ and
generates the reflected GP}\label{Fig1}
\end{center}
\end{figure}

For simplicity and symmetry we assume that graphene is free-standing. Let us represent the magnetic field of the incident GP in the region $x<0$ as follows (we use the notation $H\equiv H_y$)
\begin{equation}\label{w1}
\begin{split}
&H_i(x,z) = \pm e^{ik_{p}x\pm ik_{pz}z},\\
  \end{split}
\end{equation}
where $+$ and $-$ correspond to the regions $z>0$ and $z<0$ respectively and
\begin{equation}\label{w2}
\begin{split}
k_{p} = g\sqrt{1-1/\alpha^2}, \q k_{pz} = -g/\alpha,
  \end{split}
\end{equation}
with $g=\omega/c$ being the free-space wavevector and $\alpha=2\pi\sigma/c$ being the normalized conductivity of graphene. Using the Maxwell equations, $z$-component of the electric field reads
\begin{equation}\label{w3}
\begin{split}
E_{zi}(x,z) = \mp\frac{k_p}{g} e^{ik_{p}x\pm ik_{pz}z}.
  \end{split}
\end{equation}

We will seek for the reflected GP in the following form:
\begin{equation}\label{w4}
\begin{split}
&H_r(x,z) = \pm r e^{- ik_{p}x\pm ik_{pz}z},\\
&E_{zr}(x,z) = \pm\frac{k_p}{g} r e^{-ik_{p}x\pm ik_{pz}z},
  \end{split}
\end{equation}
where we have introduce the reflection coefficient $r$. Then the total fields in the region $x<0$ read
\begin{equation}\label{w5}
\begin{split}
&H_<(x,z) = H_i(x,z)+H_r(x,z) = \\
&\pm (e^{ik_{p}x}+re^{- ik_{p}x}) e^{\pm ik_{pz}z},\\
&E_{z<}(x,z) = E_{zi}(x,z)+E_{zr}(x,z)= \\
&\mp(e^{ik_{p}x}-re^{- ik_{p}x})\frac{k_p}{g} e^{\pm ik_{pz}z}.
  \end{split}
\end{equation}

Now let us represent the field in the region $x>0$ in the form of the plane waves expansion
\begin{equation}\label{w6}
\begin{split}
&H_>(x,z) = \int_{-\infty}^\infty dk_z h(k_z) e^{ik_{x}x+ ik_{z}z},\\
&E_{z>}(x,z) = -\int_{-\infty}^\infty dk_z h(k_z)\frac{k_x}{g} e^{ik_{x}x+ ik_{z}z},
  \end{split}
\end{equation}
where $k_x = k_x(k_z)= \sqrt{g^2-k_z^2}$

The boundary conditions along the line $x=0$ read
\begin{equation}\label{w7}
\begin{split}
&H_<(0,z) = H_>(0,z),\\
&E_{z<}(0,z) = E_{z>}(0,z).
  \end{split}
\end{equation}
Using \eqref{w6} and \eqref{w5} we have
\begin{equation}\label{w8}
\begin{split}
&(1+r) \psi(z) = \int_{-\infty}^\infty dk_z h(k_z) e^{ik_{z}z},\\
&(1-r)\frac{k_p}{g} \psi(z) = \int_{-\infty}^\infty dk_z h(k_z)\frac{k_x}{g} e^{ik_{z}z},
  \end{split}
\end{equation}
where $\psi(z)$ is the GP profile function
\begin{equation}\label{w9}
\begin{split}
\psi(z) = \pm e^{\pm ik_{pz}z}.
  \end{split}
\end{equation}

At $x=0$ the z-component of the electric field must be continuous, so we project the second line of \eqref{w9}
by $\frac{e^{-ik_{z}z}}{2\pi}$. However, the magnetic field does not need to be continuous, due to the
existence of surface currents, so we project the first line of \eqref{w9} by $\psi(z)$. After integration, we obtain:
\begin{equation}\label{w10}
\begin{split}
&h(k_z)\frac{k_x}{g} =(1-r)\frac{k_p}{g}\frac{1}{2\pi} \int_{-\infty}^\infty dz\psi(z)e^{-ik_{z}z},\\
&(1+r) \int_{-\infty}^\infty dz\psi^2(z) = \int_{-\infty}^\infty dk_zdz h(k_z)\psi(z) e^{ik_{z}z},
  \end{split}
\end{equation}
Substituting $h(k_z)$ from the first line into the second one, we find the linear equation for $r$:
\begin{equation}\label{w11}
\begin{split}
&(1+r) \int_{-\infty}^\infty dz\psi^2(z) = \\
&(1-r)\frac{k_p}{g }\frac{1}{2\pi}\int_{-\infty}^\infty dk_zdzdz' \frac{g}{k_x}\psi(z)\psi(z')e^{ik_{z}(z-z')}.
  \end{split}
\end{equation}
The integrals in $z$, $z'$ are trivially performed
\begin{equation}\label{w12}
\begin{split}
 &\int_{-\infty}^\infty dz\psi^2(z) = \frac{1}{|k_{zp}|},\\
 &\int_{-\infty}^\infty dzdz' \psi(z)\psi(z')e^{ik_{z}(z-z')} = \frac{4k_{z}^2}{(|k_{zp}|^2+k_z^2)^2},
  \end{split}
\end{equation}
Then \eqref{w11} becomes
\begin{equation}\label{w13}
\begin{split}
(1+r)\frac{1}{|k_{zp}|} = (1-r)\frac{2k_p}{\pi}\int_{-\infty}^\infty dk_z \frac{k_{z}^2}{k_x(|k_{zp}|^2+k_z^2)^2}.
  \end{split}
\end{equation}

\begin{figure}[h!]
\begin{center}
\includegraphics[width=8cm]{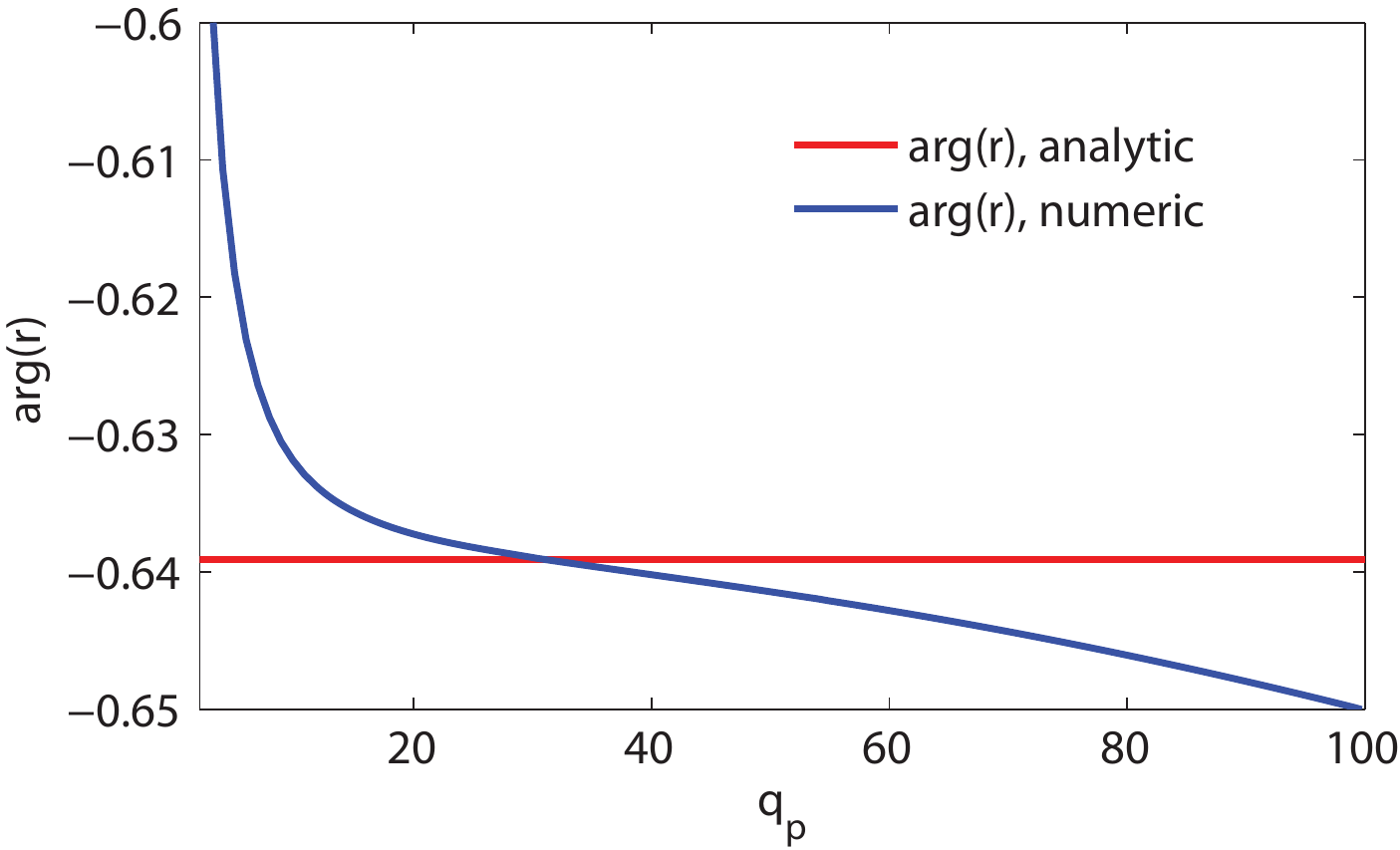}\\
\caption{The calculated phase of the reflection coefficient for GP at different values of momentum $q_p$. Both analytic (red line) and numeric (blue curve) results are shown.}\label{FigS2}
\end{center}
\end{figure}

From here we find the reflection coefficient explicitly
\begin{equation}\label{w14}
\begin{split}
r = \frac{a-1}{a+1},
  \end{split}
\end{equation}
where
\begin{equation}\label{w15}
\begin{split}
a = \frac{2|q_{zp}|q_p}{\pi} \int_{-\infty}^\infty dq_z \frac{q_{z}^2}{q_x(|q_{zp}|^2+q_z^2)^2}.
  \end{split}
\end{equation}
In the last expression the dimensionless components of the wavevectors read
\begin{equation}\label{w16}
\begin{split}
q_{zp} = -\frac{1}{\alpha}, \q q_p = \sqrt{1-\frac{1}{\alpha^2}}, \q q_x = \sqrt{1-q_z^2}.
  \end{split}
\end{equation}

The results of the numeric calculations for the reflection coefficient phase as a function of the GP momentum are shown in Fig.2 (blue curve). The integral in \eqref{w15} has been calculated by using Cauchy theorem to avoid the branch point singularity $q_x=0$ ($q_z = 1$).

The integral \eqref{w15} can be strongly simplified if we take into account that the major contribution comes from large wavevectors, i.e from those with $q_z\gg1$. In this case $q$ in the denominator of the integrand can be replaced by $q\simeq iq_z$. Then the integration is analytic and the result reads simply
\begin{equation}\label{w17}
\begin{split}
a = -2i/\pi,
  \end{split}
\end{equation}
so that the result is independent upon the GP wavevector and thus properties of graphene. According to Eqs.~\eqref{w17},\eqref{w14} the reflection phase reads
\begin{equation}\label{w18}
\begin{split}
\arg(r) = \mathrm{arctan}(-4 \pi / (4 + \pi^{2})) \approx - 0.64 \pi.
  \end{split}
\end{equation}
This result is shown in Fig.~2 (red line) and is in an excellent agreement with the numerical simulations.

\section{Inhomogeneously doped graphene}

\subsection{Local reflection coefficient}

\begin{figure}[h!]
\begin{center}
\includegraphics[width=8cm]{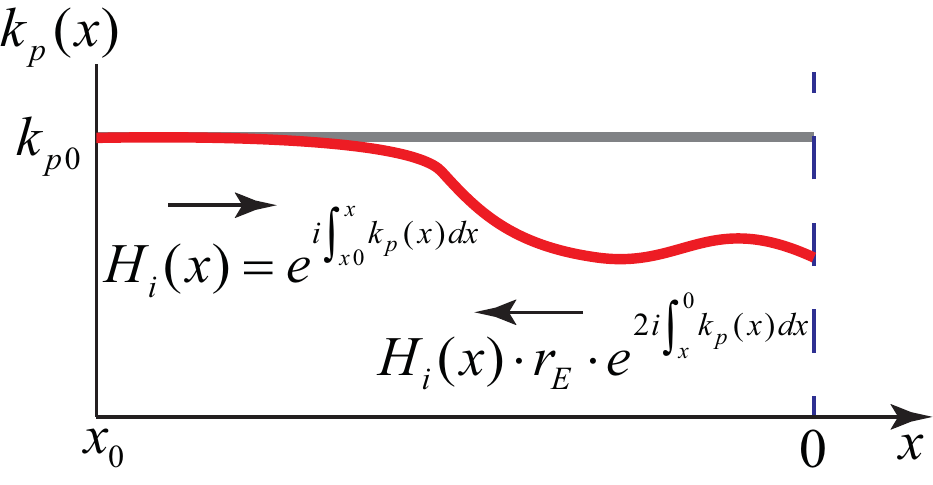}\\
\caption{The schematic of the geometrical optics approximation for graphene plasmon reflection from the edge in the inhomogeneously doped graphene.}\label{FigS3}
\end{center}
\end{figure}

Let us assume now that the conductivity of graphene is a function of distance, $\sigma=\sigma(x)$, see Fig.~\ref{FigS3}. Then if the variation is smooth enough, the plasmon propagation can be described by the geometrical optics. Namely, introducing the local plasmon wavevector $k_p(x)\propto1/\sigma(x)$, one can represent the magnetic field of the incident plasmon propagating towards the edge as follows (we will omit the $z$-dependency assuming that we are staying infinitesimally close to the graphene face, for example at $z=0^+$):
\begin{equation}\label{i1}
\begin{split}
H_i(x) = e^{i\int_{x_0}^x dx' k_p(x')},
  \end{split}
\end{equation}
where $x_0$ is a point where the phase of the magnetic field is zero. The incident plasmon propagates toward the edge (located at the position $x=0$), experiences the reflection with the reflection coefficient $r_E$ and propagates back. Then taking into account the progressive phase accumulation, the resulting reflected plasmonic field $H_r(x)$ can be written as
\begin{equation}\label{i2}
\begin{split}
H_r(x) = H_i(x)r_Ee^{2i\int_{x}^0 dx' k_p(x')}.
  \end{split}
\end{equation}
The total magnetic field $H_{tot}(x)$ presents the sum of the incident and reflected fields $H_{tot}(x)=H_i(x)+H_r(x)$, so that
\begin{equation}\label{i3}
\begin{split}
H_{tot}(x) = e^{i\int_{x_0}^x dx' k_p(x')} + r_Ee^{i\int_{x_0}^x dx' k_p(x')+2i\int_{x}^0 dx' k_p(x')}.
  \end{split}
\end{equation}
Let us rewrite this equation in another useful form:
\begin{equation}\label{i4}
\begin{split}
H_{tot}(x) = e^{i\varphi_0+i\Delta\varphi_i(x)}\left[e^{ik_{p0}x} + r(x)e^{-ik_{p0}x}\right],
  \end{split}
\end{equation}
where we have introduced the initial phase $\varphi_0=-k_{p0}x_0$, the phase incursion for the incident field
\begin{equation}\label{i5}
\begin{split}
\Delta\varphi_i(x)= \int_{x_0}^x dx' \Delta k_p(x')
  \end{split}
\end{equation}
with $\Delta k_p(x) = k_p(x)-k_{p0}$; and a local reflection coefficient
\begin{equation}\label{i6}
\begin{split}
r(x) = r_E e^{i\Delta\varphi(x)}, \q \Delta\varphi(x) = 2i\int_{x}^0 dx' \Delta k_p(x').
  \end{split}
\end{equation}
Once the total field is known (for instance from the full wave simulations), both the local reflection coefficient $r(x)$ and reflection coefficient for the edge $r_E$ can be extracted from Eqs.~\eqref{i4}-\eqref{i6}.

\subsection{The shape of the conductivity profile}

For the calculations of the reflection phase in case of the inhomogeneous profile of the conductivity, we will use the inverse square root type dependency in the vicinity of the edge $\propto 1/\sqrt{x}$. In order to avoid the singularity at $x=0$, we displace the singular point to positive values of $x$, so that the profile has the following form
\begin{equation}\label{p1}
\begin{split}
\sigma(x) = \sigma_0F(x),
  \end{split}
\end{equation}
where $\sigma_0$ is the conductivity far away from the perturbation (or "background" conductivity), $\sigma_0 = \sigma(-\infty)$ under the assumption that $F(-\infty)=1$. The function $F(x)$ reads
\begin{equation}\label{p1}
\begin{split}
F(x) =
\left\{
\begin{matrix}
1, \q x<x_0,\\
\frac{f(x)}{f(x_0)}, \q  x_0\leq x\leq0,\\
0, \q x>0,\\
\end{matrix}
\right.
  \end{split}
\end{equation}
with
\begin{equation}\label{p2}
\begin{split}
f(x) = 1+\frac{a}{\sqrt{|x-\delta|}},
  \end{split}
\end{equation}
where $\delta$ and $a$ are the parameters controlling the maximal value of the conductivity at the position $x=0$ (at the edge) and the growth, while $x_0$ presents the distance from the edge where the conductivity saturates to the constant value. In the calculations for the paper we take $x_0=-1\, \mu$m and $\delta=10$nm.

Notice that we do not follow any special model for the shape of the conductivity, but take the above profile for the proof of principle.

\section{The ribbon-ribbon interaction}

Fig. 4 shows the effect of the distance between the ribbons in the array. One can clearly see that for a larger separation (period) $L$
the lowest-frequency resonance ($n=0$) in the relative transmission is much closer to the curve predicted by a model (the white line).

\begin{figure}[h!]
\begin{center}
\includegraphics[width=8cm]{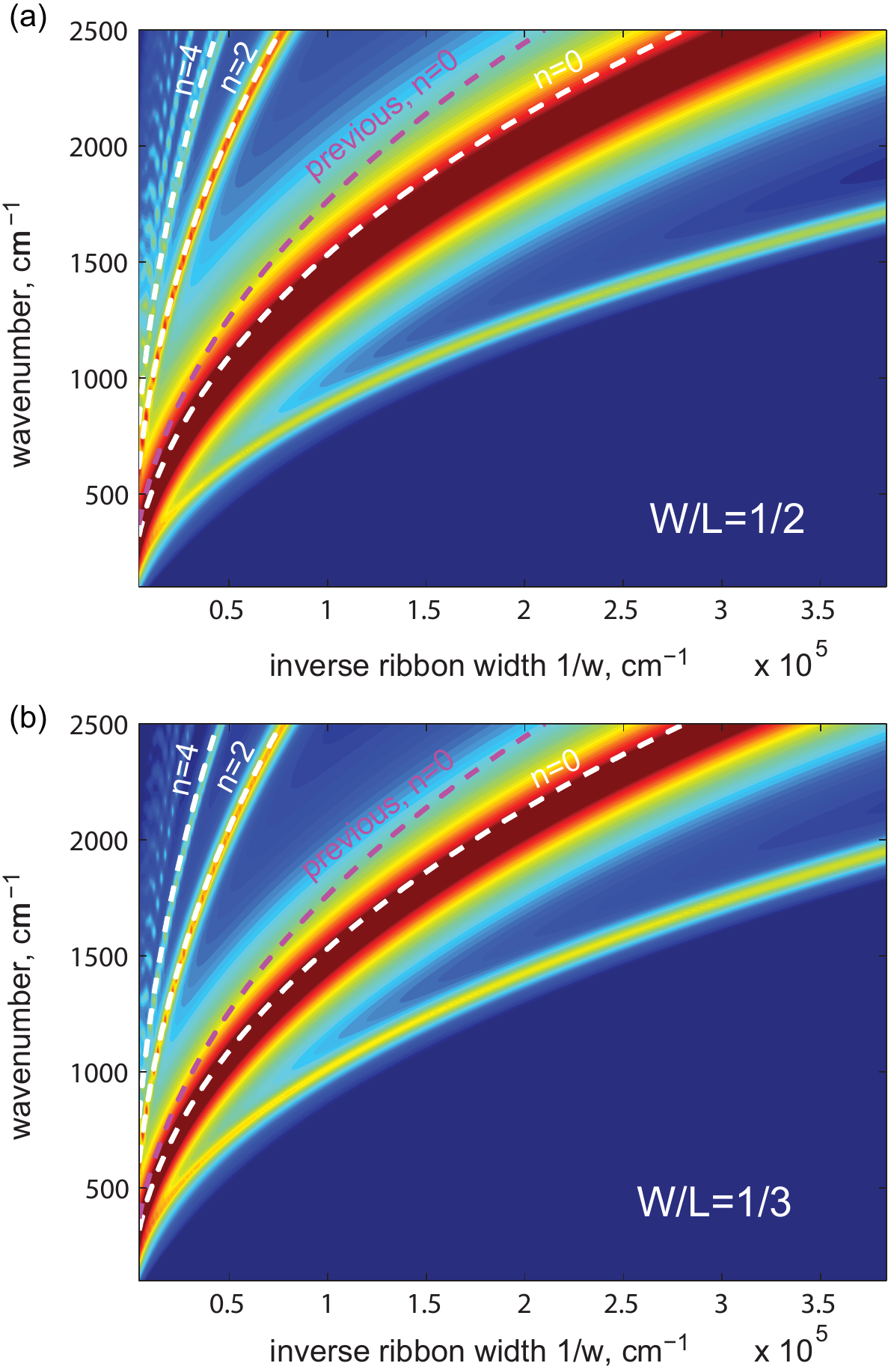}\\
\caption{Relative transmission $\delta T = |1-T/T_0|$ (with respect to the one for the graphene-free substrate) through the array of free-standing graphene ribbons as a function of the inverse ribbon width and wavenumber. The discontinuous curves numbered by ``0,2,...'' correspond to the modes given by the model (Eq.~(6) in the manuscript). The discontinuous curve marked as ``previous, n=0'' corresponds to $\Phi_R=-\pi$ and $n=0$. The parameters used for graphene conductivity: temperature $300$ K, Fermi level $0.55$ eV, relaxation time of the charge carriers is $0.1$ ps. (a) The period equal to twice the ribbon width, $L=2W$. (b) $L=3W$. }\label{FigS4}
\end{center}
\end{figure}


\begin{thebibliography}{99}
\bibitem{COMSOL} A commercial software \textsc{comsol} has been used. Graphene has been simulated as a two-dimensional conductivity in the boundary conditions.
\end{thebibliography}
\end{document}